\documentclass[prd,letterpaper,onecolumn,amsmath,amsfonts,amssymb,nofootinbib]{revtex4}
\pdfoutput=1
\usepackage {amsmath}
\usepackage[dvips]{graphicx}
\usepackage{feynmp}
\usepackage{amssymb}
\newcommand{\be}{\begin{equation}}
\newcommand{\ee}{\end{equation}}
\newcommand{\bear}{\begin{eqnarray}}
\newcommand{\eear}{\end{eqnarray}}

\newcommand{\lapproxeq}{\lower .7ex\hbox{$\;\stackrel{\textstyle
<}{\sim}\;$}}
\newcommand{\gapproxeq}{\lower .7ex\hbox{$\;\stackrel{\textstyle
>}{\sim}\;$}}
\newcommand{\stackdown}[2]{\lower 1.4ex\hbox{$\;\stackrel{\textstyle{#1}}
{\scriptstyle{#2}}\;$}}
\newcommand{\beq}{\begin{equation}}
\newcommand{\eeq}{\end{equation}}

\newcommand{\ba}{\begin{eqnarray}}
\newcommand{\ea}{\end{eqnarray}}

\newcommand{\bea}{\begin{eqnarray}}
\newcommand{\eea}{\end{eqnarray}}

\makeatletter
\def\slash{\@ifnextchar[{\fmsl@sh}{\fmsl@sh[0mu]}}
\def\fmsl@sh[#1]#2{%
  \mathchoice
    {\@fmsl@sh\displaystyle{#1}{#2}}%
    {\@fmsl@sh\textstyle{#1}{#2}}%
    {\@fmsl@sh\scriptstyle{#1}{#2}}%
    {\@fmsl@sh\scriptscriptstyle{#1}{#2}}}
\def\@fmsl@sh#1#2#3{\m@th\ooalign{$\hfil#1\mkern#2/\hfil$\crcr$#1#3$}}
\makeatother
\usepackage{color}

\definecolor{orange}{rgb}{0.9,0.2,0}
\definecolor{brown}{rgb}{0.7,0.3,0.2}
\definecolor{fuxia}{rgb}{1,0,1}
\definecolor{skyblue}{rgb}{0,0.1,0.9}
\definecolor{violetred}{rgb}{0.8,0.13,0.56}
\definecolor{deeppink}{rgb}{1.00,0.08,0.5}
\definecolor{pink}{rgb}{1.00,0.75,0.80}
\definecolor{orchid}{rgb}{0.85,0.44,0.84}
\definecolor{lightpink}{rgb}{1.00,0.71,0.76}
\definecolor{bluish}{rgb}{0,0.6,0.8}
\begin{document}
\title{Inflationary behavior of $R^2$ gravity in a conformal framework}
\author{A. B.~\ Lahanas}
\email{alahanas@phys.uoa.gr}
\affiliation{University of Athens, Physics Department,
Nuclear and Particle Physics Section,
GR--157 71  Athens, Greece}
\author{K. Tamvakis}
\email{tamvakis@uoi.gr}
\affiliation{University of Ioannina,
Theoretical Physics Department,
Section of Theoretical Physics,
GR--451 10  Ioannina, Greece}
\vspace*{2cm}
\begin{abstract}
Models of inflation are tightly constrained by the PLANCK satellite data. Among them, Starobinsky's model with an exponential type potential seems to be challenged by the recent BICEP2 results. The model is based on the existence of $\, ~ R^2$ terms in the Einstein-Hilbert action, which  have their origin in the conformal anomaly. Conformal (or Weyl) gravitational theories are relevant when matter fields  become effectively massless;  i.e.  their masses are negligible in comparison with the spacetime curvature. These theories may include other, additional scalar fields. We show that their presence under general conditions does not destabilize the inflationary behavior encountered in the Starobinsky model, although the issue of the exact quantitative agreement with existing data, like the tensor to scalar ratio, rests on the choice of parameters.
\end{abstract}
\maketitle

\section{Introduction }

The first round of data on CMB from the PLANCK satellite experiment \cite{PLANCK} is in comfortable agreement with the predictions of the Starobinsky model of inflation. Nevertheless, very recently, preliminary data from the BICEP2 experiment \cite{BICEP} challenge this fact showing a disagreement in the predicted tensor to scalar ratio. As a result, the model has received a lot of attention
\cite{KEHAGIAS}.
Starobinsky's model \cite{STAR} is based on the existence of $\, R^2$ terms that could arise due to the conformal anomaly of a classical conformal theory. In the high curvature regime masses can be neglected and quantum corrections can be approximated by the quantum fluctuations of massless conformally invariant fields. Part of these corrections to the energy-momentum tensor can be summed into a local $R^2$ term in the effective action\cite{VIL}. The resulting action reads\footnote{We use a metric with signature $(-1,\,+1,\,+1,\,+1)$.}
\begin{equation}
{\cal{S}}\,=\,\int\,d^4x\,\sqrt{-g}\,\left\{\,\frac{1}{2}R\,+\,\frac{\alpha}{2}R^2\,\right\}\,.\label{STARO}\end{equation}
Since the quantum corrections that give rise to $R^2$ are logarithmically divergent and counterterms are needed, the parameter $\alpha$ is arbitrary. Note that these are the only acceptable quartic terms in the action, since terms $R_{\mu\nu\rho\sigma}R^{\mu\nu\rho\sigma}$ are expressed in terms of $R^2$ and $R_{\mu\nu}R^{\mu\nu}$ through the Gauss-Bonnet identity, and a term $\beta\,R_{\mu\nu}R^{\mu\nu}$ introduces a spin-2 {\textit{poltergeist}} (ghost) with mass $\beta^{-1/2}$ that decouples only in the $\beta\rightarrow 0$ limit  \cite{STELLE,Barth:1983hb,Cecotti:1987sa}. 
This issue will be discussed later. 

The action ({\ref{STARO}}), besides the standard massless spin-2 graviton, contains an additional scalar degree of freedom which can become manifest if we introduce an auxiliary scalar field variable $\Phi$ and write the action in the classically equivalent form \cite{Whitt:1984pd}
\begin{equation}{\cal{S}}\,=\,\int\,d^4x\,\sqrt{-g}\,\left\{\,\frac{1}{2}\left(1+2\Phi\right) \,R\,-\frac{\Phi^2}{2\alpha}\,\right\}\,.{\label{SCALAR}}
\end{equation}
By Weyl-rescaling the metric according to $g_{\mu\nu}\,=\,\Lambda\,\overline{g}_{\mu\nu}$,
we transform the action into the form
$$
{\cal{S}}\,=\,\int\,d^4x\,\sqrt{-\overline{g}}\,\Lambda^2\,\left\{\,\frac{1}{2}\frac{(1+2\Phi)}{\Lambda}\overline{R}\,-\frac{3}{4}  \, (1+2\Phi) \,\frac{(\overline{\nabla}\Lambda)^2}{\Lambda^3}\,-\frac{\Phi^2}{2\alpha}\,\right\}$$
or, taking $\Lambda\,=\,(1+2\Phi)^{-1}$, into
\begin{equation}\int\,d^4x\,\sqrt{-\overline{g}}\,\left\{\,\frac{1}{2}\overline{R}\,-\frac{1}{2}(\overline{\nabla}\sigma)^2\,-\frac{1}{8\alpha}\left(1\,-e^{-\sqrt{\frac{2}{3}}\sigma}\,\right)^2\,\right\}.{\label{POT-0}}
\end{equation}
For the last step we introduced the canonically normalized  field $\sigma\equiv\sqrt{3/2}\ln(1+2\Phi)$.
The scalar potential of the Starobinsky model in the form ({\ref{POT-0}}) shows clearly an inflationary behavior.

In the present article we reconsider $R^2$ gravity in the more general framework of conformally invariant theories and study its behavior with respect to inflation. We start with a Weyl invariant action of gravitation and a scalar field, incorporating the breaking induced by the conformal anomaly in an $R^2$ term with an arbitrary coefficient. First, we study a version of this theory distinct from the Starobinsky model, which as it stands does not lead to a satisfactory slow-roll inflation, although, additional scalar fields, conformally coupled, could modify that. Next, we consider a version that includes the Starobinsky model and shares a generic inflationary behavior. We show that, under general conditions, additional scalar fields, conformally coupled to this model, sustain this behavior. 
Nevertheless, the issue of the exact quantitative agreement with existing data, like the tensor to scalar ratio, rests on the choice of parameters.

\section{General Conformally Invariant Framework}

Consider the following Weyl invariant action 
\begin{equation} {\cal{S}}\,=\,\int\,d^4x\,\sqrt{-g}\,\left(\,\alpha_0'\,(C_{\mu\nu\rho\sigma})^2\,
-\frac{s}{2}\left(\,\frac{X^2}{6}\,R\,+\,\,(\nabla X)^2\,\right)\,-\lambda\,X^4\,+\,\Delta{\cal{L}}\,\right)
{\label{CONFO}}
\end{equation}
written in terms of the {\textit{Weyl tensor}}\footnote{ $$C_{\rho\sigma\mu\nu}\,\equiv\,R_{\rho\sigma\mu\nu}\,-\frac{2}{(D-2)}\left(g_{\rho\mu}R_{\nu\sigma}+g_{\rho\nu}R_{\mu\sigma}-g_{\sigma\mu}R_{\nu\rho}-g_{\sigma\nu}R_{\mu\rho}\right)+\frac{2R}{(D-1)(D-2)}\left(g_{\rho\mu}g_{\nu\sigma}+g_{\rho\nu}g_{\mu\sigma}\right)$$} $C_{\mu\nu\rho\sigma}$ and a scalar field $X$. The parameter $s$ is restricted by conformal invariance to the values $s=0,\pm 1$. The parameter $\alpha_0'$ is dimensionless. Finally, $\Delta{\cal{L}}$ stands for conformally invariant interactions of $X$ with additional fields. For $s=1$, $X$ is a canonical field but the gravitational Einstein term does not have a positive sign. In contrast, for $s=-1$ the field $X$ is a ghost and has to be fixed (conformal gauge fixing) but the Einstein term has the correct sign.
{{{The Lagrangian}}}
({\ref{CONFO}}) is invariant under the following {\textit{Weyl or conformal transformations}}
\begin{equation}\left\{\begin{array}{l}
g_{\mu\nu}\rightarrow\,\Lambda(x)\,g_{\mu\nu}\\
\,\\
X\rightarrow\,\Lambda^{-1/2}(x)\,X\end{array}\right.\,{\label{TRANSF}}\end{equation}
for any $\Lambda(x)$.

Applying the Gauss-Bonnet theorem, the Weyl term takes the form
\begin{equation}
\alpha_0'\,\int\,d^4x\,\sqrt{-g}\,C_{\mu\nu\rho\sigma}^2\,=\,\frac{\alpha_0'}{2}\int\,\sqrt{-g}\,\left(\,R_{\mu\nu}R^{\mu\nu}\,-\frac{1}{3}R^2\,\right)\,+\,\dots\,
{\label{C}}
\end{equation}
where the ellipsis denotes a topological term (Euler number) that does not contribute to the equations of motion. This term is conformally invariant. Nevertheless, as we mentioned in the Introduction, conformally coupled matter can generate at the one-loop quantum level a  term of the form 
\begin{equation}
\frac{\alpha_0}{2}\int\,d^4x\,\sqrt{-g}\,R^2
\label{R2}
\end{equation}
which breaks conformal invariance and is induced as a result of the trace anomaly. The anomaly generated effective action includes, in addition, nonlocal terms \cite{Riegert:1984kt} that can be made local by the introduction of auxiliary fields \cite{ANTOMOTTO}. For an extensive discussion on this issue see  \cite{Fabris:1998vq,Hawking:2000bb,Fabris:2000gz}.

The variation of the combined quadratic action terms ({\ref{C}}) and ({\ref{R2}}) gives
\begin{equation}
\frac{2}{\sqrt{-g}}\frac{\delta{\cal{S}}}{\delta g^{\mu\nu}}\,=\,\alpha_0'\left(\, W_{\mu\nu}^{(2)}\,-\frac{1}{3}W_{\mu\nu}^{(1)}\,\right)\,+\,\alpha_0\,W_{\mu\nu}^{(1)}\,,\end{equation}
where the tensors $W_{\mu\nu}^{(2)}$ and $W_{\mu\nu}^{(1)}$ stem from the variation of $R_{\mu\nu}R^{\mu\nu}$ and $R^2$ respectively\footnote{These tensors are \cite{BIR}
$$\begin{array}{l}
W_{\mu\nu}^{(1)}\equiv\,\frac{1}{\sqrt{-g}}\frac{\delta}{\delta g^{\mu\nu}}\int\,d^4x\,\sqrt{-g}\,R^2\,=\,2\nabla_{\mu}\nabla_{\nu}R\,-2g_{\mu\nu}\Box R\,-\frac{1}{2}g_{\mu\nu}R^2\,+\,2RR_{\mu\nu}\\
\,\\
W_{\mu\nu}^{(2)}\,\equiv\,\frac{1}{\sqrt{-g}}\frac{\delta}{\delta g^{\mu\nu}}\int\,d^4x\,\sqrt{-g}R_{\rho\sigma}R^{\rho\sigma}\,=\,2\nabla_{\nu}\nabla_{\rho}R_{\mu}^{\,\,\rho}\,-\Box R\,-\frac{1}{2}g_{\mu\nu}\Box R\,+\,2R_{\mu}^{\,\,\rho}R_{\rho\nu}\,-\frac{1}{2}g_{\mu\nu}R^{\rho\sigma}R_{\rho\sigma}
\end{array}$$}. The first term is identically traceless thanks to the Bianchi identity, as expected, since it arises from a conformally invariant $C^2$ term of the action. In addition, it vanishes for Friedmann-Roberston-Walker (FRW) geometries, yielding no contribution to the equations of motion. 

From the previous discussion it becomes evident that as far as the equations of motion are concerned the  terms given by Eq. (\ref{C}) 
are Weyl invariant and do not contribute to the equations of motion when we consider conformally flat geometries,  and in particular FRW  cosmologies.  Therefore only the term (\ref{R2}), quadratic in the Ricci scalar $R$, plays an essential role in the dynamics and will be  kept in the action. Then,  as we shall see, after appropriate Weyl rescalings, and a suitable gauge-fixing of the Weyl symmetry,  the gravitational part  receives the  well-known Einstein form. 
However the presence of the terms (\ref{C}) in the action is essential in order to study the quantum behavior of gravity in the UV regime.  Taking these terms into account the gravitational part of the Lagrangian density in the Einstein frame, denoted by  barred quantities, takes on the form
\bea
\frac{1}{2} \, \bar{R} + \frac{\alpha^\prime_0}{2} \, \left(  {\bar{R}}_{\mu \nu}^2 - \frac{1}{3} \, \bar{R}^2 \right) 
\eea
In it the last two terms are Weyl invariant and  have no effect when studying FRW cosmologies, as already discussed. This Lagrangian was studied by Stelle ( see second reference in \cite{STELLE} ) and it is a renormalizable gravity which however includes ghost states, invalidating therefore the unitarity of the theory. 
In general higher derivative gravities are better behave in the UV but they suffer, in general, by the presence of negative norm states ( ghosts ) \cite{GHOST,BISWAS1,BISWAS2}.  
This subtle issue has been analyzed in the literature, where the most general gravity action was considered which involves terms up to quadratic in the Riemmann tensor $\, R_{\mu \nu k \lambda} \,$, in an attempt to find a resolution towards building theories of gravity that do not pose UV problems and are ghost free.  It has been shown that the completion of the gravitational action by higher derivative nonlocal operators may render a ghost free theory \cite{BISWAS1, BISWAS2}.  
Nevertheless, the study of inflation in such a framework is beyond the scope of this paper.

Thus, in what follows we shall restrict ourselves to the action
\begin{equation} {\cal{S}}\,=\,\int\,d^4x\,\sqrt{-g}\,\left(\,\frac{\alpha}{2}R^2\,
-\frac{s}{2}\left(\,\frac{X^2}{6}\,R\,+\,\,(\nabla X)^2\,\right)\,-\lambda\,X^4\,+\,\right)\,+\,\Delta{\cal{S}},\end{equation}where $\alpha$ is the renormalized value of the corresponding dimensionless parameter. This action, apart from the $R^2$ term which signals the breaking of conformal symmetry at the quantum level, is invariant under the conformal transformations ({\ref{TRANSF}}). $\Delta{\cal{S}}(g_{\mu\nu},\,X,\,\sigma)$ contains conformally invariant interactions with extra fields denoted collectively with $\sigma$.
In what follows we shall consider first the action without the presence of extra fields. 
Two distinct cases exist depending on the sign of $s$. 

\section{The case ${s=+1}$} 
In the case $s=+1$
the kinetic term for the $X$ field has the correct sign 
and the action is
\begin{equation}{\cal{S}}_0\,=\,\int\,d^4x\,\sqrt{-g}\,\left(\,\frac{\alpha}{2}R^2\,
-\frac{1}{12}\,X^2\,R\,-\frac{1}{2}(\nabla X)^2\,-\lambda\,X^4\,\right)\end{equation}
or, introducing the auxiliary field $\Phi$,
\begin{equation}{\cal{S}}_0\,=\,\int\,d^4x\,\sqrt{-g}\,\left(\,\left(\Phi\,-\frac{X^2}{12}\right)R\,-\frac{\Phi^2}{2\alpha}\,-\frac{1}{2}(\nabla X)^2\,-\lambda\,X^4\,\right)\,.\end{equation}
Despite the superficial resemblance with the Starobinsky model ({\ref{SCALAR}}), the action ${\cal{S}}_0$ is clearly different since, for the chosen sign of $s$, the term linear in curvature
, $ -\frac{1}{12}\,X^2\, R \,$, 
has the opposite sign of the standard Einstein term. Thus, the inflationary behavior driven by the conformal anomaly encountered in the Starobinsky model and embodied in the potential ({\ref{POT-0}) is not necessarily expected. 

Next, we perform a {\textit{Weyl rescaling}} of the metric accompanied by a field redefinition
\be g_{\mu\nu}\,=\,\Lambda\,\overline{g}_{\mu\nu},\,\,X\,=\,\Lambda^{-1/2}\,\overline{X}\,.{\label{RESC}}\ee
{{{  Taking $\Lambda$ as}}}
\footnote{$\overline{X}^2\,=\,X^2/\left(2\Phi\,-\frac{X^2}{6}\right)$}
$$\Lambda\left(\Phi\,-\frac{X^2}{12}\right)\,=\,\frac{1}{2}\,\Longrightarrow\,\Lambda\,=\,\frac{1}{2\Phi}\left(1+\frac{\overline{X}^2}{6}\right)\,,$$
we obtain
\begin{equation}
{\cal{S}}_0\,=\,\int\,dx^4\,\sqrt{-\overline{g}}\,\left\{\,\frac{1}{2}\overline{R}\,-\frac{3}{4}\left(1+\frac{\overline{X}^2}{6}\right)\left(\frac{\overline{\nabla}\Phi}{\Phi}\right)^2\,-\frac{1}{2}\frac{(\overline{\nabla}\overline{X})^2}{\left(1+\frac{\overline{X}^2}{6}\right)}-\lambda\overline{X}^4\,-\frac{1}{8\alpha}\left(1+\frac{\overline{X}^2}{6}\right)^2\,\right\}\,.
\end{equation}
Introducing the field variables
\begin{equation}
\overline{X}\,=\,\sqrt{6}\,\sinh\psi,\,\,\,\,\,
\Phi\,=\,e^{2\phi},\,
\end{equation}
we can write the action in the form
\begin{equation}{\cal{S}}_0\,=\,\int\,dx^4\,\sqrt{-\overline{g}}\,\left\{\,\frac{1}{2}\overline{R}\,-3\cosh^2\psi\,(\overline{\nabla}\phi)^2\,-3(\overline{\nabla}\psi)^2\,-36\lambda\sinh^4\psi\,-\frac{1}{8\alpha}\cosh^4\psi\,\right\}{\label{ACT}}\,.
\end{equation}
The resulting scalar field equations of motion, dropping the overline bars for simplicity of notation, are
\begin{equation}
\begin{array}{l}
\Box\phi\,+\,2\tanh\psi(\nabla^{\mu}\psi)(\nabla_{\mu}\phi)\,=\,0\\
\,\\
\Box\psi\,=\,\cosh\psi\sinh\psi\,\left(\,(\nabla\phi)^2\,+\,24\lambda\,\sinh^2\psi\,+\,\frac{1}{12\alpha}\cosh^2\psi\,\right)\end{array}
\end{equation}
For a flat  FRW metric $ds^2\,=\,-dt^2\,+\,a^2(t)d\vec{x}^2$, these equations take on the form
\begin{equation}
\begin{array}{l}
\ddot{\phi}+3H\dot{\phi}+2\dot{\psi}\dot{\phi}\tanh\psi\,=\,0\\
\,\\
\ddot{\psi}+3H\dot{\psi}=\,-\sinh\psi\cosh\psi\,\left(\,-\dot{\phi}^2\,+\,24\lambda\sinh^2\psi\,+\,\frac{1}{12\alpha}\cosh^2\psi\,\right)\end{array}{\label{EOMS+}}
\end{equation}
where 
$H\equiv \dot{a} /{a}$. 
The corresponding Friedmann equation reads
\begin{equation}
3H^2\,=\,\rho\,=\,3\cosh^2\psi\,\dot{\phi}^2+3\dot{\psi}^2\,+\,36\lambda\sinh^4\psi+\frac{1}{8\alpha}\cosh^4\psi\,.
\end{equation}

Although the general solution of the coupled system of equations ({\ref{EOMS+}}) is hard to obtain, a partial class of solutions with $\dot{\phi}=0$ reduces the system to just one equation for $\psi$, namely  
\begin{equation}\ddot{\psi}\,+\,3H\dot{\psi}\,=\,-V'(\psi)/6{\label{EOM1}}\end{equation}
with 
\begin{equation}V(\psi)\,=\,36\lambda\sinh^4\psi+\frac{1}{8\alpha}\cosh^4\psi\,.{\label{POT}}\end{equation}

The potential in ({\ref{POT}}) cannot drive inflation. It is convenient to introduce  {{{ the parameter }}}$A\equiv\,1/(288\alpha\lambda)$ and write it as
$V(\psi)=36  \lambda\left( \, A  \cosh^4  \psi \, +  \sinh^4  \psi \right)$. 
The minimum of the potential is at $\, \psi_0 = 0$ and its minimum value is $\, V_{min} = 1 / 8 \alpha $. The quantity $\, F \equiv -V'(\psi)/6  $ on the rhs of ({\ref{EOM1}}) vanishes at $\, \psi_0 = 0 $. Furthermore, its first derivative at this point is $\, F^\prime_0 = - 24 A  \lambda < 0 \,$, i.e. is negative. Therefore, the point $\, \psi_0 = 0 $ is an attracting fixed point and, whatever its initial value, $\psi$ will be attracted towards  $\, \psi_0 = 0 $. Nevertheless, the potential has no flat directions to guarantee that the {\textit{slow-roll conditions}} for inflation can be met. It is flat in the vicinity of the fixed point but the  attraction to it  is not slow. Taking care 
of the noncanonical normalization of the field $\psi$, we write down the slow-roll parameter $\epsilon$
\be
\varepsilon \, \equiv \, - \, \frac{\dot H}{H^2} \, = \, \frac{1}{12} \,  {\left(  \frac{V^\prime}{V} \right)}^2
 \, = \, \frac{4}{3} \, t^2 \, {\left(    \frac{A+t^2}{A+t^4} \right)}^2
 \ee 
 with $ t \equiv \tanh \psi$.
This vanishes at the fixed point mentioned above and there is,  therefore, a range around $\, \psi_0 = 0 $ for which $\epsilon $ is small as required for inflation. However the approach to this point is rather fast since the other slow-roll condition can never be met. In fact the parameter $\eta$, with the aforementioned normalization, is
\be
\eta \, = \, \frac{1}{6} \,  \left(  \frac{V^{''}}{V} \right)
 \, = \, \frac{2}{3} \, \frac{ A ( 4 \, s^4 + 5 \, s^2 + 1 ) + 3 \, s^4 + 4 s^2  }{ A c^4 + s^4 }\,,\ee 
 with $s\equiv\sinh\psi$ and $c\equiv\cosh\psi$.
At $\, \psi_0 = 0 $ 
{{{
this reaches its minimum value  $\, \eta = 2/3  $ which is already large, implying that he attraction to the fixed point takes place with large acceleration. }}}Therefore, without going into the details, we are convinced that the model cannot sustain inflation.
Thus, although the general framework of conformal invariance seems to be the right framework to investigate inflation driven by the conformal anomaly, for the choice $s=+1$, at least in the minimal case of one field $X$, no suitable inflationary behavior is sustained.

We shall not proceed to analyze the possible inflationary behavior induced by the presence of the extra fields, since this will be entirely attributed to the extra fields and it is bound to depend on details of their action. Instead, we shall proceed to consider the more interesting $s=-1$ case which includes the standard Starobinsky model and starts up having a generic inflationary behavior in the minimal case.

\section{THE CASE $s=-1$}
Let's go back to the original action and take the opposite sign of the parameter $s=-1$. The action expressed in terms of the auxiliary field $\Phi$ is
\begin{equation}
{\cal{S}}_0\,=\,\int\,d^4x\,\sqrt{-g}\,\left\{\,\left(\Phi\,+\,\frac{X^2}{12}\right)R\,-\frac{\Phi^2}{2\alpha}\,+\,\frac{1}{2}(\nabla X)^2\,-\lambda X^4\,\right\}\,.
\end{equation}
Now the Einstein-like term linear in the curvature has the right sign but the field $X$ is a ghost having the wrong sign in its kinetic term. A common procedure is to fix the field $X$ by a {\textit{conformal gauge condition}}. As such, the condition $X=\sqrt{6}$, reducing the linear coupling to the curvature into its Einstein value, is often used \cite{LINDE}. 
In the absence of $R^2$ terms, that break conformal invariance, this gauge choice can be interpreted as spontaneous breaking of the conformal symmetry \cite{LINDE,KALLOSH2}. 
Performing a Weyl rescaling ({\ref{RESC}}) accompanied by a field redefinition, we go to the Einstein frame where the action takes the form
\begin{equation}
{\cal{S}}_0\,=\,\int\,d^4x\,\sqrt{-\overline{g}}\,\left\{\,\frac{1}{2}\overline{R}\,-\frac{3}{4}\left(\,1\,-\frac{\overline{X}^2}{6}\right)\left(\frac{\overline{\nabla}\Phi}{\Phi}\right)^2\,+\,\frac{1}{2}\frac{(\overline{\nabla}\overline{X})^2}{\left(1-\frac{\overline{X}^2}{6}\right)}\,-\lambda\overline{X}^4\,-\frac{1}{8\alpha}\left(1-\frac{\overline{X}^2}{6}\right)^2\,\right\}\,.
\end{equation}
The gauge condition on the scalar field reads
\begin{equation}X\,=\,\sqrt{6}\,\,\Longrightarrow\,\overline{X}\,=\,\frac{\sqrt{6}}{\sqrt{1+2\Phi}}\,.{\label{GAUGE}}\end{equation}
Inserting this condition into the action, it takes the form
\begin{equation} {\cal{S}}_0\,=\,\int\,d^4x\,\sqrt{-\overline{g}}\,\left\{\,\frac{\overline{R}}{2}\,-3\frac{(\overline{\nabla}\Phi)^2}{(1+2\Phi)^2}\,-\frac{36\lambda}{(1+2\Phi)^2}\,-\frac{1}{8\alpha}\left(\frac{2\Phi}{1+2\Phi}\right)^2\,\right\}\end{equation}
or, in terms of the field $\phi\,=\,\sqrt{\frac{3}{2}}\,\ln(1+2\Phi)$, 
\begin{equation}{\cal{S}}_0\,=\,\int\,d^4x\,\sqrt{-\overline{g}}\,\left\{\,\frac{1}{2}\overline{R}\,-\frac{1}{2}(\overline{\nabla}\phi)^2\,-\frac{1}{8\alpha}\left(\,1\,-e^{-\sqrt{\frac{2}{3}}\phi}\right)^2\,-36\lambda\,e^{-2\sqrt{\frac{2}{3}}\phi}\,\right\}\,.{\label{EQUA}}\end{equation}
This is exactly the Starobinsky model in scalar language with an extra $\lambda$ self-interaction term which does not have any drastic effect on its general inflationary behavior. Thus, for the choice $s=-1$ inflation is a property of the minimal conformal theory without the presence of extra fields. Nevertheless, the question of whether the detailed slow-roll inflationary behavior is in quantitative agreement with existing data is open. {{{Moreover}}}}, the question whether this general inflationary behavior persists in the presence of extra conformally coupled scalar fields is not an empty one.
The role of additional fields that couple in a conformally invariant manner cannot be excluded in general. 
For instance additional  fields may exist in effective  gravity theories, having their origin in string or superstring theories, and at very high energies their couplings are conformal invariant since their masses can be neglected.  
The  role of additional scalars coupled in a conformally invariant manner in the context of cosmological inflation models,  in a different context, has been also considered in other works, see for instance \cite{LINDE,KALLOSH2,BARS} . 
In the following section we will take up and investigate the simple case that  an additional  scalar field is present. Evidently more involved scenarios are possible. 
\subsection{Extra fields}
Let's consider an additional scalar field coupled in the action through the extra term
\be\,
\Delta{\cal{S}}\,=\,\int\,d^4x\,\sqrt{-\overline{g}}\,\overline{X}^4\,{\cal{P}}(\sigma,\,K)\,{\label{PPP}}
\ee
where $K$ stands for
\be 
K\,\equiv\,\frac{(\overline{\nabla}\sigma)^2}{2\overline{X}^2}\,.
\ee
We assume that the field $\sigma$ has zero Weyl weight, being 
therefore 
invariant under the conformal transformations ({\ref{TRANSF}}). Then, $K$ is Weyl invariant too and so is the full action $\Delta{\cal{S}}$ having the same form in both Jordan and Einstein frames that are connected by these transformations. Such Lagrangians have been considered in the context of K-inflation\cite{armend1,armend2} and also in higher derivative theories in which the vacuum is ghost free (ghost-condensate vacua) 
\cite{ArkaniHamed:2003uz,ArkaniHamed:2003uy,Weinberg:2008hq,Koehn:2012te,ARKANI2}
{\footnote{
The Lagrangian above for the $\sigma$-field is the Weyl-invariant generalization of a similar action where $\bar{X}=constant$, occurring in ordinary (non-Weyl) gravity models  which employ the shift symmetry $\sigma\rightarrow\,\sigma\,+\,const.$ 
 \cite{ArkaniHamed:2003uz,ArkaniHamed:2003uy,Weinberg:2008hq,Koehn:2012te,ARKANI2}. 
}}
. 
Thus, we add ({\ref{PPP}})
to the minimal action ${\cal{S}}_0$ ({\ref{ACT}}). Introducing the field variables
\begin{equation}
1+2\Phi\,=\,e^{\sqrt{\frac{2}{3}}\phi},\,\,\,\,\,\overline{X}\,=\,\sqrt{6}\,\sin\psi\,,
\end{equation}
the action takes the form
\begin{equation} {\cal{S}}\,=\, \int\,d^4x\,\sqrt{-\overline{g}}\,\left\{\frac{1}{2}\overline{R}\,-\frac{1}{2}(\overline{\nabla}\phi)^2\left(\frac{\cos^2\psi}{\left(1-e^{-\sqrt{\frac{2}{3}}\phi}\right)^2}\right)\,+\,3(\overline{\nabla}\psi)^2\,-V(\psi)\,\right\}\,+\,\Delta{\cal{S}}\,.
\end{equation}
On the other hand, the gauge condition ({\ref{GAUGE}}) becomes $\overline{X}\,=\,\sqrt{6}\,e^{- \frac{1}{2} \sqrt{\frac{2}{3}}\phi}$ and, enforcing it on the action, we obtain
\begin{equation}
{\cal{S}}\,=\,\int\,d^4x\,\sqrt{-\overline{g}}\,\left\{\,\frac{1}{2}\overline{R}\,-\frac{1}{2}(\overline{\nabla}\phi)^2-V(\phi)\,+\,\overline{X}^4{\cal{P}}(K,\sigma)\,\right\}\,.
\end{equation}
In the equation above $V(\phi)\,=\,\frac{1}{8\alpha}\left(\,1\,-e^{-\sqrt{\frac{2}{3}}\phi}\right)^2\,+\,36\lambda\,e^{-2\sqrt{\frac{2}{3}}\phi}$. 

We shall assume, that the theory is invariant under constant translations $\sigma\rightarrow\,\sigma\,+\,const.$ and, therefore, ${\cal{P}}$ is only a function of $K$. Such solutions are analogous to the so-called {{``ghost condensate''}} solutions related to the spontaneous breaking of Lorentz invariance
as has been already discussed 
\cite{ArkaniHamed:2003uz,ArkaniHamed:2003uy,Weinberg:2008hq,Koehn:2012te,ARKANI2}.

The corresponding equations of motion are

\begin{equation}
\begin{array}{l}
\frac{1}{\sqrt{-\overline{g}}}\partial_{\mu}\left(\sqrt{-\overline{g}}\,\overline{X}^2\,{\cal{P}}_K\,\partial^{\mu}\sigma\,\right)\,=\,0\\
\,\\
\Box\phi\,=\,V'(\phi)\,+\,\sqrt{\frac{2}{3}}\,\overline{X}^4\left(2{\cal{P}}\,-K{\cal{P}}_K\,\right)
\end{array}
\end{equation}
with ${\cal{P}}_K\equiv\frac{\partial{\cal{P}}}{\partial K}$. In a FRW geometry these equations are
\begin{equation}
\begin{array}{l}
\frac{d}{dt}\left(\,a^3\,\overline{X}^2\,{\cal{P}}_K\,\dot{\sigma}\,\right)\,=\,0\\
\,\\
\ddot{\phi}\,+\,3H\,\dot{\phi}\,=\,-V'(\phi)\,-\sqrt{\frac{2}{3}}\,\overline{X}^4\left(2{\cal{P}}\,-K{\cal{P}}_K\,\right)
\end{array}{\label{OMEGA}}
\end{equation}
The corresponding energy and momentum densities are
\begin{equation}
\begin{array}{l}
\rho\,=\,\frac{1}{2}\dot{\phi}^2\,+\,V(\phi)\,-\overline{X}^2{\cal{P}}_K\dot{\sigma}^2-{\cal{P}}\overline{X}^4\\
\,\\
p\,=\,\frac{1}{2}\dot{\phi}^2\,-V(\phi)\,+\,{\cal{P}}\overline{X}^4
\end{array}
\label{rhopre}
\end{equation}
{{{
When $\dot{\phi}^2 << V(\phi) $ the {\textit{null energy condition}}  $\rho + p > 0$ holds as long as ${\cal{P}}_K < 0$ and  for  ${\cal{P}}_K=0$ we have $\rho + p=0$ or, equivalently, a barotropic index $w=-1$ analogous to a cosmological constant.
}}}

Note that the set of equations ({\ref{OMEGA}}) depends on the variables $\phi$ and $\omega\,\equiv\,\dot{\sigma}$ and can be cast in the form
\begin{equation}
\begin{array}{l}
\frac{d}{dt}\left(\,a^3\,\frac{\partial V_{eff}}{\partial\omega}\,\right)\,=\,0\\
\,\\
\ddot{\phi}\,+\,3H\,\dot{\phi}\,=\,-\frac{\partial V_{eff}}{\partial\phi}
\end{array}
\label{eq36}
\end{equation}
with
\begin{equation}
V_{eff}\,=\,V(\phi)\,-\overline{X}^4\,{\cal{P}}\,,
\end{equation}
while the energy-density and pressure, given by (\ref{rhopre}), can be written as 
\begin{equation}
\rho\,=\,\frac{1}{2}\dot{\phi}^2\,+\,V_{eff}\,-\omega\frac{\partial V_{eff}}{\partial\omega},\,\,\,\,\,\,\,p = \frac{ {\dot {\phi}}^2}{ 2} -  V_{eff}
\end{equation}
Considering a velocity expansion and restricting ourselves up to a quartic velocity term
\begin{equation}
{\cal{P}}\,=\,-f_0\,K\,-\frac{g_0}{2}K^2\,,{\label{P}}
\end{equation}
with $f_0,\,g_0$ constants, {{{which is mandatory if the symmetry $ \sigma \rightarrow \sigma + const. $   is imposed }}}
{\footnote{Also, as a result of this symmetry, the scalar potential for the field $\sigma$  can be only a constant which however  can be incorporated in the quartic coupling $\lambda$. Therefore this symmetry leads to the minimal scenario, as far as the number of the free parameters describing the model are concerned. It would be nice to see, in the more general case, that the shift symmetry solution corresponds to an attractor but this is beyond the scope of this work. }}
, we obtain for the $V_{eff}$ given above
\begin{equation}
V_{eff}\,=\,V(\phi)\,-\frac{1}{2}\mu^2(\phi)\,\omega^2\,+\,\frac{g_0}{8}\omega^4\,,
\end{equation} 
with
\begin{equation}
\mu^2(\phi)\,\equiv\,f_0\,\overline{X}^2\,=\,6f_0\,e^{-\sqrt{\frac{2}{3}}\phi}\,.
\end{equation}
Note that the effective potential is now the sum of the Starobinsky potential and a {\textit{Higgs-like potential}}.

\subsection{A CLASS OF SOLUTIONS}
Looking for solutions of the above equations, we first consider the class of solutions with ${\cal{P}}_K=0$, corresponding to the minima of $V_{eff}$ with respect to $\omega$
\begin{equation}
{\cal{P}}_K\,=\,0\,\Longrightarrow\,\frac{\partial V_{eff}}{\partial\omega}\,=\,0\,.
\end{equation}
For the chosen form ({\ref{P}}) of ${\cal{P}}$ they imply
\begin{equation}
\begin{array}{l}
\omega^2\,=\,2 
\mu^2(\phi)/g_0\,\Longrightarrow\,\dot{\sigma}^2\,=\,\frac{12f_0}{g_0}e^{-\sqrt{\frac{2}{3}}\phi}\\
\,\\
\ddot{\phi}\,+\,3H\,\dot{\phi}\,=\,-18\,\xi\sqrt{\frac{2}{3}}\,e^{-2\sqrt{\frac{2}{3}}\phi}\,-\frac{1}{4\alpha}\sqrt{\frac{2}{3}}\,e^{-\sqrt{\frac{2}{3}}\phi}\\
\,\\
3H^2\,=\,\frac{1}{2} {\dot{\phi}}^2 \,+\,V(\phi) - 18 \, \frac{f_0^2}{g_0}e^{-2\sqrt{\frac{2}{3}}\phi}\,
\end{array}
{\label{EQU-1}}
\end{equation}
where we have introduced the parameter $\xi$ as 
\begin{equation}
\xi\,\equiv\,4\lambda\,+\,\frac{1}{72\alpha}\,-\frac{2f_0^2}{g_0}\,.{\label{KSI}}
\end{equation}

It is clear from ({\ref{EQU-1}}) that we have a solution with constant $\phi$, namely
\begin{equation}
\phi_0\,=\,\sqrt{\frac{3}{2}}\,\ln( \, 72\alpha\xi \,),\,\,\,\,\,\dot{\sigma}_0^2\,=\,\frac{f_0}{6g_0\alpha\xi}\,.{\label{STATIC}}
\end{equation}
The corresponding Hubble parameter is constant and it is given by the Friedmann equation as
\begin{equation}
H_0^2\,=\,\frac{1}{24\alpha}
\left(1-\frac{1}{72\alpha\xi}\right)\,.
\label{papafr}
\end{equation}
This {\textit{"static"}} solution corresponds to a minimum of $V_{eff}$, satisfying
\begin{equation}
\frac{\partial V_{eff}}{\partial\omega}\,=\,\frac{\partial V_{eff}}{\partial\phi}\,=\,0\,.
\end{equation}

Linear stability of ({\ref{STATIC}}) can be readily checked. Perturbing around this solution as
\begin{equation}
\phi\,\approx\,\sqrt{\frac{3}{2}}\,\ln( \, 72\alpha\xi \, ) \,+\,\delta\phi,\,\,\,\,\,\sigma\,\approx\,t\sqrt{\frac{f_0}{6g_0\alpha\xi}}\,+\,\delta\sigma\,,
\end{equation}
we are led to
\begin{eqnarray}
&& \delta\ddot{\phi}+3H_0 \, \delta\dot{\phi}\,+\,(432\alpha^2\xi)^{-1}\delta\phi\,=\,0\, \\
&& \delta\dot{\sigma}\,\pm\,(f_0/36\alpha g_0\xi)^{1/2}\delta\phi \, = \, 0
\end{eqnarray}
with solutions
\begin{equation}
\delta\phi\,=\,B\,e^{-\Gamma t},\,\,\,\,\,\,\,\delta\dot{\sigma}\,\approx\, \mp \frac{B}{2}\sqrt{\frac{2}{3}}\sqrt{\frac{f_0}{6g_0\alpha\xi}}\,e^{-\Gamma t}\,,
\end{equation}
with
\begin{equation}
\Gamma\,=\,\frac{3H_0}{2}\left(1\pm\sqrt{1-\frac{H_c^2}{H_0^2}}\,\right)\,\,\,\,\,\,\,\,\,\,\,\,\,\,\,\left(H_c^2\,\equiv\,\frac{1}{972 \alpha^2 \xi}\right)\,.
\end{equation}
Note that for 
$H_0^2<H_c^2\,\,\,\Longrightarrow\, \sqrt{72 \alpha\xi } <{{\frac{5}{3}}},$ the correction
$\delta \phi$ has an oscillatory factor. As for the corrections to the Hubble rate, one finds from the Friedmann equation, by a straightforward computation using the first  order result (\ref{papafr}), that 
\begin{equation}
6 H_0\,\delta H\,\approx\,
\delta \phi \left(  V^\prime(\phi_0) +  \sqrt{\frac{2}{3}} \, \frac{36 \, f_0^2}{g_0} 
e^{-2\sqrt{\frac{2}{3}}\phi_0}
 \right)
\end{equation}
The term within the bracket on the rhs of this equation is actually, up to a minus sign, the lhs of Eq.  
(\ref{EQU-1}), i.e $ \ddot{\phi_0}\,+\,3H\,\dot{\phi_0}\,$, which vanishes since $\phi_0$ is a constant. As a result 
\begin{equation}
\delta H\,=\,0\,.
\end{equation}

The solution ({\ref{STATIC}}) belongs to the restricted class satisfying $\frac{\partial V_{eff}}{\partial\omega}=0$. However this condition cannot be met for arbitrary initial conditions. In fact, the general solution of the equation of motion for $\sigma${{{, as can be seen from (\ref{eq36}),}}} is 
\begin{equation}
\frac{\partial V_{eff}}{\partial\omega}\,=\, - \frac{\cal{C}}{a^3}
\label{mineff}
\end{equation}
with ${\cal{C}}$ a constant. This is equivalent to being at a minimum of
\begin{equation}
\tilde{V}_{eff}\,=\,V_{eff}\,+\,\frac{\cal{C}}{a^3} \,\omega\,=\,-\frac{1}{2} \mu^2(\phi) \, \omega^2\,+\,\frac{g_0}{8} \,\omega^4\,+ 
\,\frac{\cal{C}}{a^3} \, \omega\,.
\end{equation}
Solutions that eventually lead to an expanding scale factor are bound to reach the minimum of $V_{eff}$, since, after an early period, the linear term will become subdominant. In this case the system will evolve according to ({\ref{STATIC}}) independently of initial conditions.

In order to study the complete solutions of (\ref{mineff}) it is convenient to rescale $\omega$ as
\begin{equation}
 \omega \, = \, {\left( \frac{2 f_0 {\bar{X}}^2}{g_0} \right)}^{1/2} \, {\Omega}
\label{re1}\,.
\end{equation}
Then, Eq.  (\ref{mineff}) takes on the form 
\begin{equation}
 {\Omega}^3 -  \, \Omega \, - f(t) \, = \, 0
\label{om2}\,,
\end{equation}
where $f(t)$ is given by 
\begin{equation} 
 f(t) \, = \, C_i \,   {\left( \bar{X} \, a   \right) }^{-3}   
\label{funt}\,.
\end{equation}
In the equation above $C_i$ is a constant, proportional to the one appearing in (\ref{mineff}). 
Thus the solution for $\Omega$ is controlled only by the function $f(t)$. This equation has only one real solution if $\, f^2(t) > \frac{4}{27} $ and three real solutions 
when $\, f^2(t) < \frac{4}{27} $. As can be easily seen from (\ref{mineff}), unless the initial velocity, when inflation starts at time $t_i$, is such that $\Omega_i$ lies in the range 
$ | \Omega_i | < 2 / \sqrt{3} $, only the first case is applicable. Thus, it seems  natural to  assume, at least for a wide range of initial conditions,  that $\, f^2(t_i) > \frac{4}{27} $ initially. 
This assumption is also supported by the fact that at $t_i$ the  value of $f^2(t_i)$ is naturally large, given the velocity $\dot\sigma_i$ and  the value of $\phi_i$, for reasonable values of $C_i$,  since the cosmic scale factor is small and also small is ${\bar X}(t_i)$  for values of $\phi_i$  in  the (almost) flat region of the potential $V(\phi)$  where  inflation starts from. Thus,  we consider 
$\, f^2(t_i) > \frac{4}{27} $ which is a valid assumption. The sign of the constant $ C_i $ in (\ref{funt}), and hence the sign of $f(t)$, depends on the sign of the initial velocity of $\sigma$, or the same of  the initial value of $\Omega$. Changing the sign of $\Omega$ simply reverts the sign of $C_i$, and hence that of $f(t)$ as can be seen from 
(\ref{om2}). Thus, the  cases $\Omega > 0$ and $\Omega <0$ are mirrors of each other and are  treated in exactly the same manner. Therefore,  
without loss of generality we  take $f(t_i) > 0$ initially, which  combined with  $\, f^2(t_i) > \frac{4}{27} $ yields   $\Omega_i > 2 / \sqrt{3}$ on account of  (\ref{om2}).  Then only one solution exists at $t_i$ but this solution evolves, since as time elapses $f(t)$ gets smaller, staying however always positive.  The reason for the decrease of $f(t)$ is that the cosmic scale factor, or more precisely the combination $\bar{X} \, a$, which enters{{{ $f(t)$ in Eq. (\ref{funt})}}},  gets larger as time increases. In fact it can be easily seen that the time derivative of $\bar{X} a$ is always positive, as $\phi$ approaches  the minimum of the potential $V(\phi)$, since the velocity of $\phi$ is negative. 
In particular using Friedmann equation it is easily found that 
$$
\frac{d \,( \bar{X} a)}{ d t} \, = \, \bar{X} \, a \, \left(    \sqrt{ \frac{\rho}{3} }  - \frac{\dot{\phi}}{\sqrt{2}} \right)\,.
$$
Thus, eventually the value of $f(t)$ becomes equal to $ \sqrt{\frac{4}{27}}$ and when this happens, at a time  $t_c$,  $\Omega(t_c)=2/\sqrt{3}$. 
The function $f(t)$ keeps decreasing, dropping below $ \sqrt{\frac{4}{27}}$,  but it stays positive,  approaching zero. 
At this point  $\Omega$ is $+1$,  which actually corresponds to the positive sign minimum of the "Higgs" potential. 
Note that although there are three real solutions when $f(t)$ drops below $ \sqrt{\frac{4}{27}}$, only one of them is continuous as function of the time at the critical point  point $t_c$.  
\begin{figure}  
\begin{center}
\includegraphics[scale=0.5]{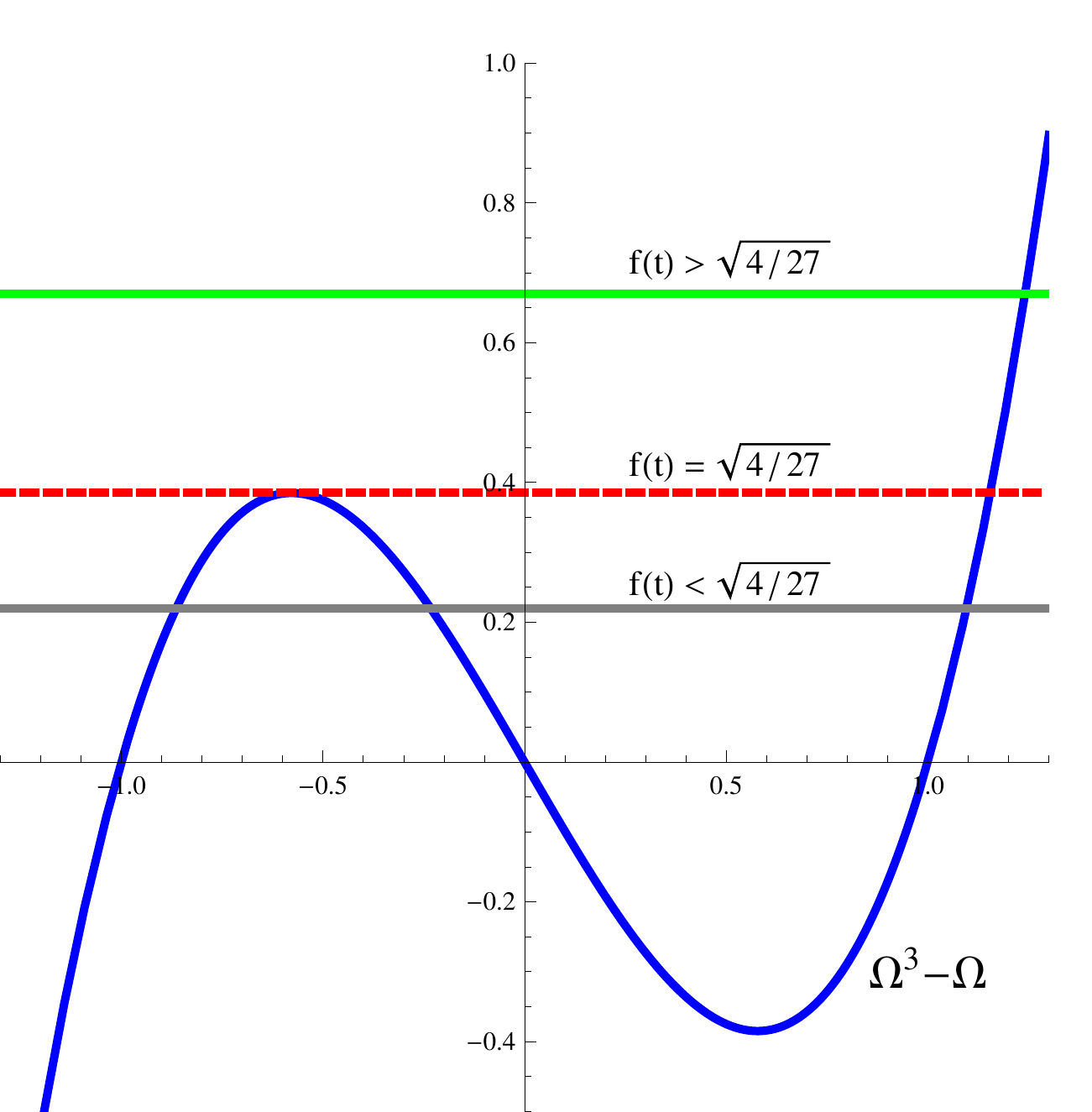}
\end{center}
\caption[]{
The form of the function $\Omega^3 -\Omega$ ( blue  line ). The function $f(t)$ for a  value larger than  $\sqrt{\frac{4}{27}}$ ( green line ), 
for values equal to $\sqrt{\frac{4}{27}}$ ( red line ), and a value smaller than  $\sqrt{\frac{4}{27}}$ ( gray line ).
}
\label{cubicfig}  
\end{figure}
The situation is depicted in Fig. \ref{cubicfig}. 
The analytic form of this solution for $\,\Omega(t)$ is given by  
\begin{equation}
{\Omega(t)} \, = \, 
\left\{
\begin{array}
{c@{\quad , \quad }c}
\frac{1}{\sqrt{3}}  \, ( \, A(t) +  {1}/{A(t)} \, ) &  if \quad  f > \sqrt{ \frac{4}{27} } \\
\frac{1}{\sqrt{3}} \,  cos \Theta(t) + sin \Theta(t) &  if \quad   \sqrt{ \frac{4}{27} } > f > 0 
\end{array}
\right.
\label{solution1}
\end{equation}
In this the functions $A(t)$ and $\Theta(t)$ are analytically given by 
\begin{equation}
 A(t) = \sqrt{1+h} - \sqrt{h} \quad , \quad \Theta(t) = \frac{1}{3} \, ArcCos ( - \sqrt{1+h} ) \quad {\text{where}} \quad h \equiv \frac{27 f^2}{4} - 1
\label{fun1}\,.
\end{equation}
Note that the function  $h$ by its definition is larger than $-1$. In particular when $ -1 < h <0 $ the range  of $f$ is  $ \sqrt{ \frac{4}{27} } > f > 0 $ and when 
$ h > 0 $ it lies in the range  $ f>\sqrt{ \frac{4}{27} } $.  
{\footnote{For the mirror case that corresponds to an opposite initial velocity $ - \Omega_i $ the function $f(t)$ is  negative, and exactly opposite to the one given by 
(\ref{solution1}) and approaches zero from below. Then $\omega$ approaches the value $-1$ corresponding to the negative sign minimum of the Higgs potential. }}

\begin{figure}  
\begin{center}
\includegraphics[scale=1.1]{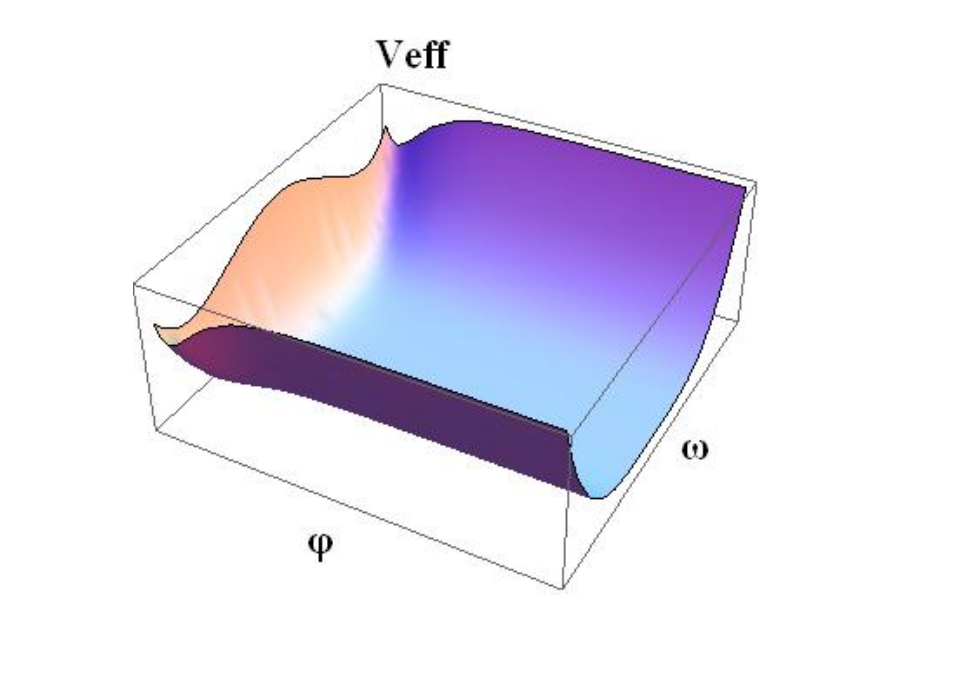}
\end{center}
\caption[]{
General shape of the two-field potential as function of $\phi$ , $\omega$ ( for explanation see main text ).
}
\label{2fieldpot}  
\end{figure}

Therefore the conclusion of this analysis is that the solution  $\Omega(t)$ tends to $+1 \, ( -1) $, provided that the initial values are such that 
$\Omega_i > 2/\sqrt{3} \, ( \Omega_i < 2/\sqrt{3} )  $.  The values $\, \Omega = \pm 1$ are actually the locations of the Higgs minima
{{{
$ \frac{\partial V_{eff}}{\partial \omega} = 0$, corresponding to ${\cal{P}}_K=0$}}}. Thus, we conclude that the solution eventually approaches either of the minima (depending on the sign of the initial $\sigma$ velocity) of the Higgs potential and then the model becomes effectively a single inflaton model governed solely by the field $\phi$. All this is supported by the generic shape of the potential shown in Fig. \ref{2fieldpot}, {{{ 
which in the $\phi$-direction has the form of the Starobinsky potential and in the $\omega$-direction has the form of a Higgs-like potential, as we have already discussed, exhibiting two symmetric  minima for  small $\phi$ values.
}}}

\subsection{SLOW-ROLL INFLATION}

Substituting the solution of the equation for $\sigma$, i.e. ${\cal{P}}_K=0$, into the other equation, we have {\textit{a single canonical field $\phi$}} moving according to
\begin{equation}\ddot{\phi}\,+\,3H\dot{\phi}\,=\,-V_{eff}'(\phi)\end{equation}
in the effective potential
\begin{equation}
V_{eff}(\phi)\,=\,\frac{1}{8\alpha}\left(\,1\,-2e^{-\sqrt{\frac{2}{3}}\phi}\,+\,L\,e^{-2\sqrt{\frac{2}{3}}\phi}\,\right)
\label{ourpot}
\end{equation}
with $L\,\equiv\,72\alpha\xi$. The minimum of the potential occurs at $\phi_0=\sqrt{\frac{3}{2}}\,\ln L$ and its value is $V_0\,=\,\frac{1}{8\alpha}\left(1-\frac{1}{L}\right)\,$. Recall that the Hubble rate $H_0$ is $3 H_0^2 = V_0$ and therefore the minimum of the potential must be positive.  Thus, the parameter $L$ should be larger than unity, $L>1$.

The corresponding slow-roll parameters are
\begin{equation}\epsilon(\phi)\,=\,\frac{1}{2}\left(\frac{V_{eff}'(\phi)}{V_{eff}(\phi)}\right)^2\,=\,\frac{4}{3}e^{-2\sqrt{\frac{2}{3}}\phi}\,\left(\frac{1-Le^{-\sqrt{\frac{2}{3}}\phi}}{1+Le^{-2\sqrt{\frac{2}{3}}\phi}\,-2e^{-\sqrt{\frac{2}{3}}\phi}}\right)^2{\label{EPSI}}\end{equation}
\begin{equation}\eta(\phi)\,=\,\frac{V_{eff}''(\phi)}{V_{eff}(\phi)}\,=\,\frac{4}{3}e^{-\sqrt{\frac{2}{3}}\phi}\left(\frac{2Le^{-\sqrt{\frac{2}{3}}\phi}\,-1}{1-2e^{-\sqrt{\frac{2}{3}}\phi}\,+\,Le^{-2\sqrt{\frac{2}{3}}\phi}}\right){\label{ETA}}\end{equation}
The slow-roll parameter $\epsilon$ is non-negative and it vanishes at the minimum of the potential exhibiting  maxima at 
$e^{\, \sqrt{ \frac{2}{3}} \, \phi_{max}} \, = \,  L \pm \sqrt{L^2 - L}$. 
The smaller of these local maxima lies below the minimum of the potential. At the other the maximum value of  $\epsilon$ is  $
\epsilon(\phi_{max}) \, = \, \left(3\,(L-1)(2L-1+2\sqrt{L^2 - L})\right)^{-1}$.
This is a decreasing function of the parameter $L$. This can be large when $L$ is close to unity. However, already for $L > 1.2$ we have that $ \epsilon(\phi_{max}) < 0.700 $. As indicative,  for $L=2$ the value of the maximum is $ \epsilon(\phi_{max}) = 057 $. 
Therefore, $\epsilon$ is in the slow-roll regime most of the time and only when $\phi$  approaches the point $\phi_{max} $ does it attain its maximum value,  which is much less than  unity when $L>2$. It is clear then that, for such values of $L$, it suffices to consider only the slow-roll parameter $\eta$ in order to study whether it might signal the exit from inflation by moving out of the slow-roll regime.
{{{ 
Note that for values of $L \geq 5$, the maximum value of $\epsilon$ is   $ \leq 0.005$, which corresponds to a{\it{ tensor to scalar ratio}}  $ r \simeq 16 \, \epsilon < 0.08$. Therefore, already from this simple analysis we see that values of $r$ in the region $r > 0.1 $ require small  values $ L < 5 $. 
}}}

If $\phi_1$ is the value for which $\eta(\phi_1)\,=\,1$ and inflation stops, 
\footnote{Note that the possibility $\eta\,=\,-1$ leads to $L<25/33$, which lies outside the allowed parameter regime of $L>1$.}
we obtain $e^{-\sqrt{\frac{2}{3}}\phi_1}=3/\left(1+\sqrt{1+15L}\right)$.
The location of $\phi_1$ with respect to the minimum $\phi_0$ of the potential gives a corresponding range for the parameter $L$. For $\phi_1>\phi_0$ we obtain
$L<7/3$ and the inflaton starts accelerating before  the minimum. The  case $L>7/3$ corresponds to $\phi_1<\phi_0$. If this were the case, the inflaton would pass beyond the minimum;  it would suffer deceleration, inverting its motion  and moving towards the minimum again, but during all this time it would still  be in the slow-roll regime. Certainly, a solution of $\eta (\phi_1)=1$ larger than the minimum is relevant, since the inflaton will pass this point as it moves towards the minimum. In contrast, a solution with $\phi_1$ smaller than the minimum does not necessarily imply that the particle reaches that point. This is supported by analyzing the real motion of $\phi$ for selected $L>7/3$ values, where $\eta$ grows as the inflaton moves towards the minimum and eventually settles there without ever becoming equal to $1$. Thus, the range of values $L>7/3$ corresponds to the inflaton reaching the minimum of the potential while still being in the slow-roll regime.

The number of $e$-folds is given by $N\,=\,\ln(a_1/a)\,=\,\int_t^{t_1}dt\,H\,$
but from the slow-roll equations $H^2\,\approx\,\frac{1}{3}V,\,\,\,3H\dot{\phi}\,\approx\,-V'$, we have $H\,\approx\,-\frac{V}{V'}\dot{\phi}$ and
$$N(\phi)\,\approx\,-\int_{\phi}^{\phi_1}\,d\phi\,\frac{V}{V'}\,.$$
For our potential, given by Eq.  (\ref{ourpot}),  we have
\begin{equation}
N\,=\,\frac{3}{4}\left\{\,\frac{1}{x}\,-\frac{1}{x_1}\,-(L-2)\ln (x/x_1)\,-(L-1)\ln\left(\frac{1-Lx_1}{1-Lx}\right)\,\right\} {\label{NFOLD}}
\end{equation}
with $x=e^{-\sqrt{\frac{2}{3}}\phi}$. 
{{{
In ({\ref{NFOLD}}) with $\phi$, corresponding to $x$,  we denote the field value at which the desired number of $e$-foldings is achieved,
while $\phi_1$ signals the end of inflation period and it is not necessarily connected with the departure of $\eta$ from the slow-roll regime. So $\phi_1>\phi_0$ is assumed, while the full range of values for $L$ is considered accessible. Note that $\phi\,>\phi_1>\phi_0$ implies $x<x_1<x_0=1/L$. Thus, we have $xL<1$. This does not necessarily imply  that $L$ is small}}}.

{{{
The number of $e$-folds ({\ref{NFOLD}}) can be considered as dependent on $L$, $z=Lx$ and $z_1=Lx$ ($z<z_1<1$). As we move $z$ away from $z_1$ towards smaller values, $N(z)$ increases rapidly especially when $z_1$ is very close to unity. 
 Although the numerical value of $N$ depends also on $z_1$, no appreciable change seems to occur as we move from the characteristic value taken $z_1=0.9$ down to $z_1=0.5$. However the exact value of $z_1$ does matter when it is taken to be very close to unity.  }}}

\begin{figure}  
\begin{center}
\includegraphics[scale=0.6]{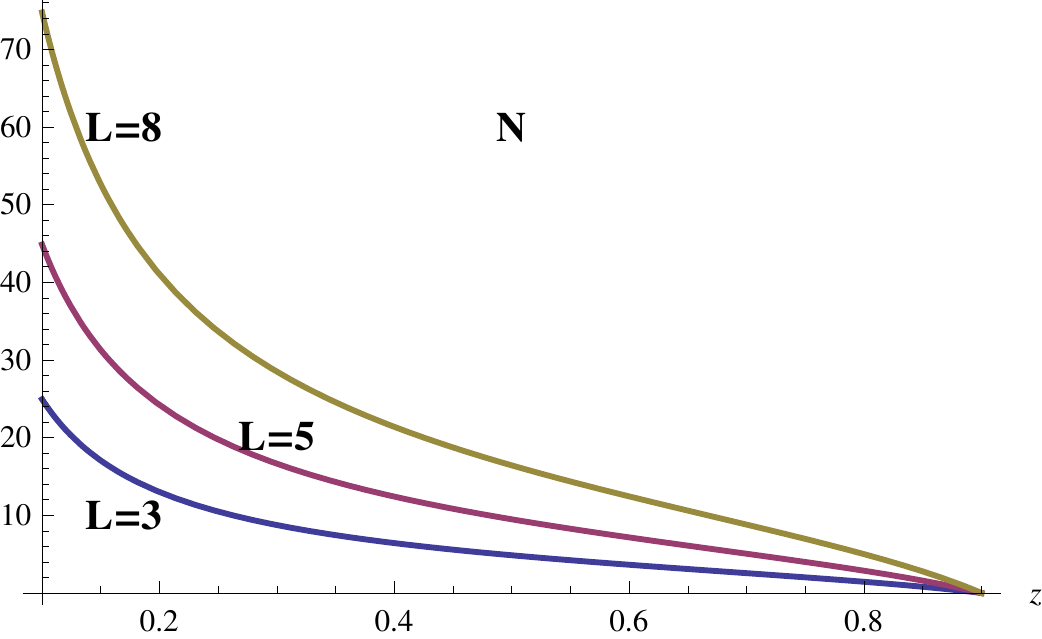}
\hspace*{4mm}
\includegraphics[scale=0.6]{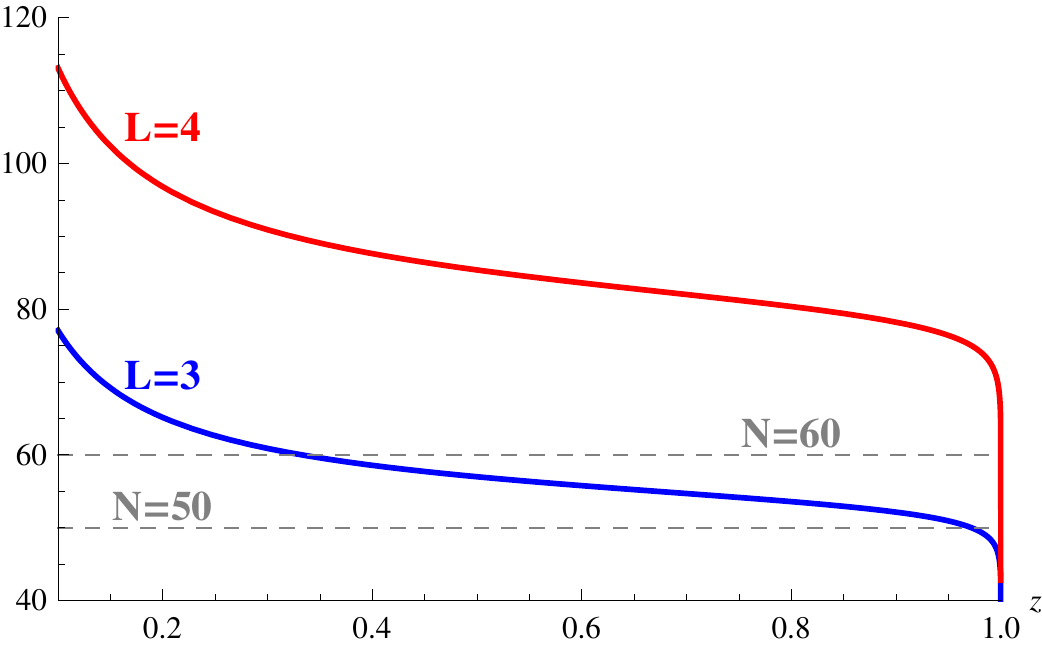}
\end{center}
\caption[]{The number of $e$-foldings, $N$,  as a function of $z$ for values  $L = 3,5  $ and $ L = 8$ ( left panel ) and $z_1 = 0.90$.  On the right panel $N$ is displayed,  when $z_1$ is almost unity, for two representative values $L=3$ and $L=4$. The lines ( dashed )  $ N = 50$ and $ N = 60 $ are shown. 
}
\label{efoldings}
\end{figure}

{{{
In Fig. \ref{efoldings} we have plotted $N(z)$ as function of $z$ for various values of the parameter $L$.
Solutions corresponding to a  point where $N(z)$ crosses a horizontal line, with values in the range $N \approx\,$ 50-60 , yield the appropriate field values to determine the spectral index $n_s$  and the tensor to scalar ratio $r$.  If this point is $z=C$ or $x=C/L$, the corresponding slow-roll parameter $\epsilon$, taken from ({\ref{EPSI}}), is 
}}}
\begin{equation}
\epsilon\,=\,\frac{4}{3}\frac{C^2(1-C)^2}{ {(L-2C+C^2) }^2 }\,.
\end{equation}  
This is small if $C$ is close to unity. As a function of $L$, $\epsilon$ becomes smaller for large values of $L$
being inversely proportional to $L^2$. It is evident from this that low  $L$ values are preferred in conjunction with values of $C$ that are not close to unity.
{{{
In the left panel of this figure three representative low $L$ cases are shown, $L=3,5,8$  with $z_1$ chosen to be $0.9$. No crossing is obtained with $L \leq 5$ and therefore in this case, as we have already discussed,  the $\epsilon$ turns out to be small entailing to small values for the ratio $r$ as well. In the right panel the value of $z_1$ has been taken very close to unity, as it should be, since the end of inflation occurs just before we reach the minimum of the potential ( when $ L > 7/3$) and two cases $ L =3$ and $ L = 4 $ are displayed. It is clearly seen that for values of $L \simeq 3$ crossing of the  $ N \simeq 60$ line occurs for values of $z \sim 0.4$, i.e. for values $ C \sim 0.4$, yielding a large value for $\epsilon$. Note also that a deviation from the value $L = 3$ towards larger values will move $C$ towards unity yielding small $\epsilon$, while if $L$ gets  smaller ($ L <2$) no crossing with the desired number of $e$-foldings, $ N \approx\,$ 50-60 ,  is obtained. 
}}}

\begin{figure}  
\begin{center}
\includegraphics[scale=0.55]{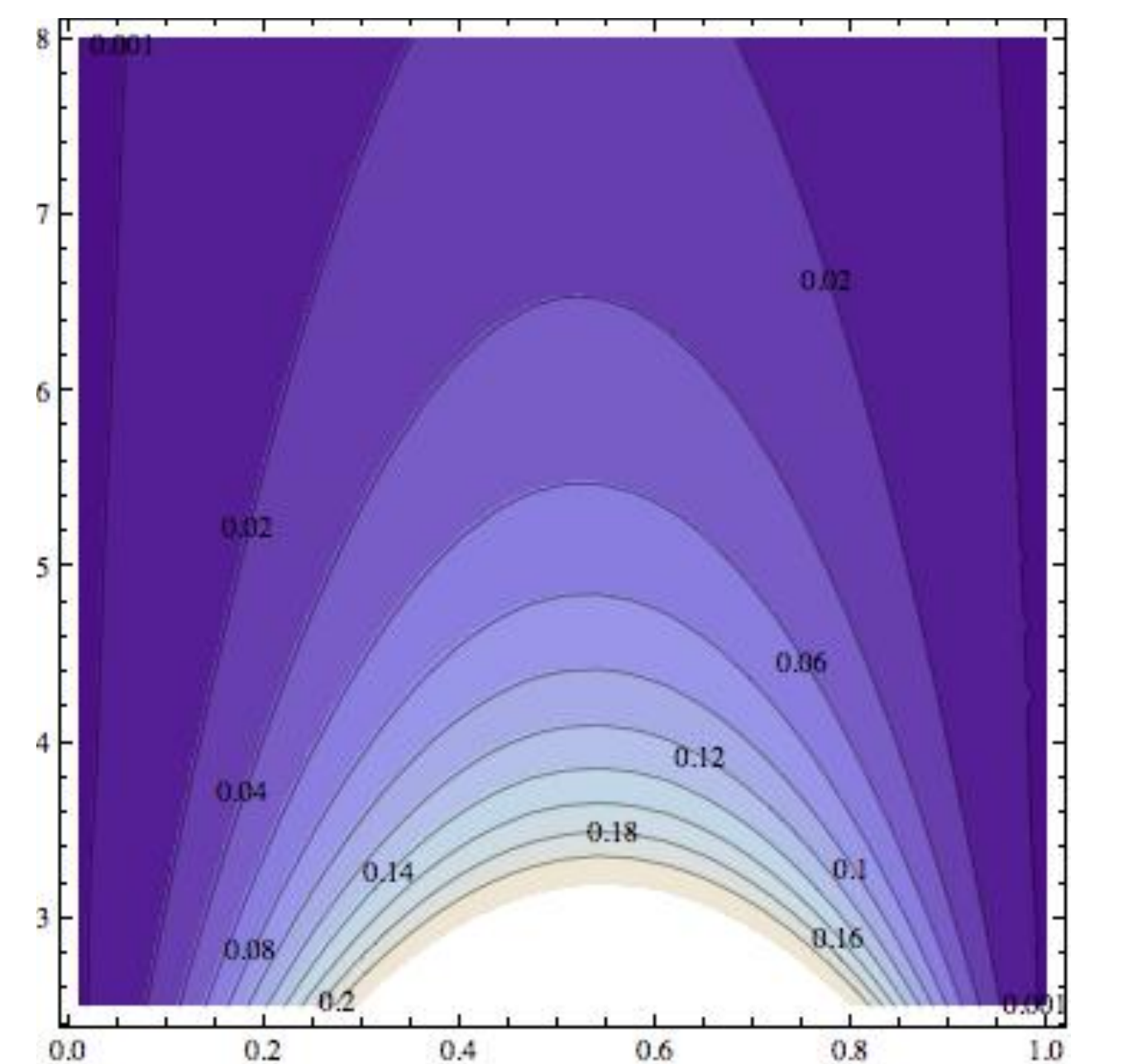}
\includegraphics[scale=0.55]{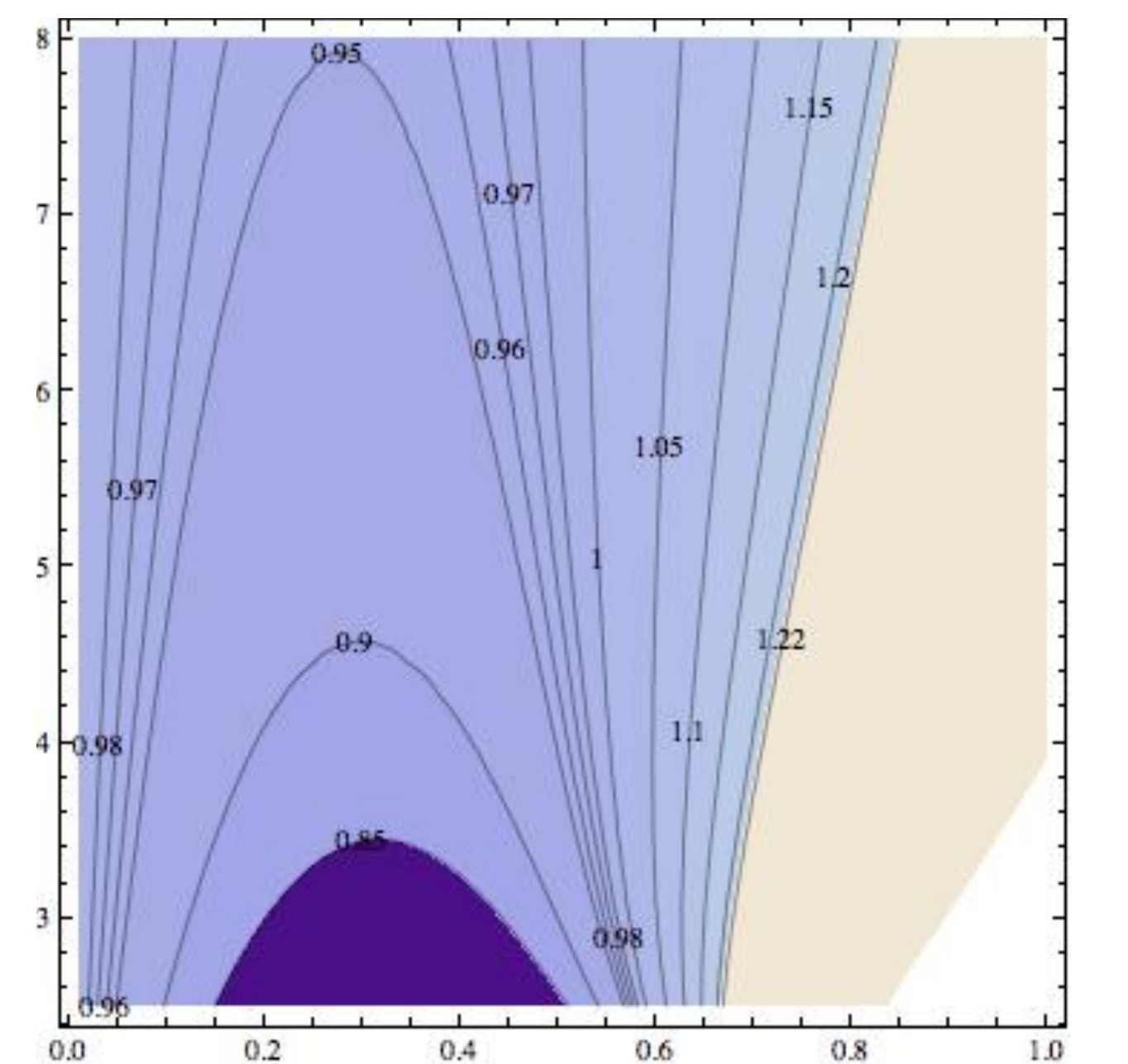}
\end{center}
\caption[]{$r$-contours ( left panel ) and $n_s$-contours (right panel). The horizontal axes are the values of  $z$ in the range 
$z = $ 0.0-1.0 . The vertical axis are the values of the parameter $L$ in the range $L = $ 2.5-8. 
}
\label{contours}
 \end{figure}

Apart from the validity of the slow-roll conditions, ensured by the smallness of $\epsilon$ and $\eta$, given by ({\ref{EPSI}}) and ({\ref{ETA}}), the viability of the model and its compatibility with the data require the consideration of the {\textit{tensor to scalar ratio}} $r$ and the spectral index $n_s$ : 
\begin{equation}
r\,\approx\,16\,\epsilon,\,\,\,\,n_s\,\approx\,1-6\epsilon+2\eta\,.
\end{equation}
In the left panel of Fig.  \ref{contours}, where we have values of $L$ in the vertical axis versus values of $z$ in the horizontal axis, we show contour lines of $r$. Although there are points with 
$r > 0.1$, for values of $L$ smaller than $4$, this should be contrasted with the corresponding values for $N$ and $n_s$. In the right panel we show the corresponding contours for $n_s$. The region favoring values  $ r > 0.1 $ and $n_s \simeq 0.96$ is located in the patch $ L < 3.5 $ and $ z= $ 0.55-0.60. The model is in agreement with both BICEP2 and PLANCK data provided the number of the $e$-foldings in this area is in the right ballpark $N = $ 50-60. Note however that in order to obtain agreement with all data $z_1$ has to be taken very close to unity. This might seem as a tuning of the model, although one would expect that the end of inflation should occur quite close to the minimum of the potential. 

\begin{figure}  
\begin{center}
\includegraphics[scale=0.6]{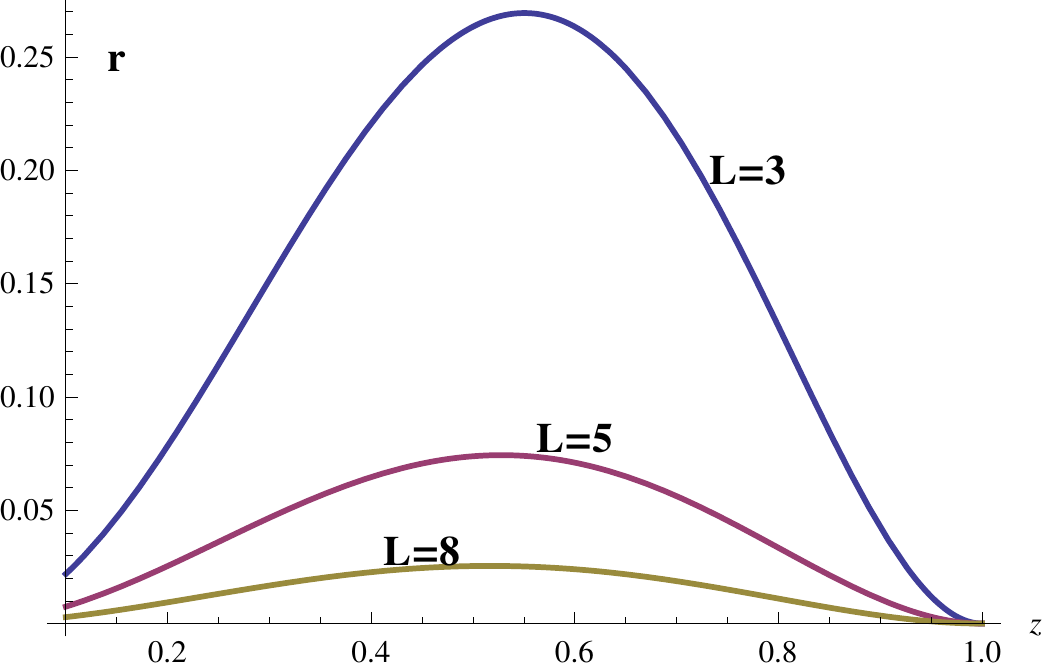}
\hspace*{15mm}
\includegraphics[scale=0.6]{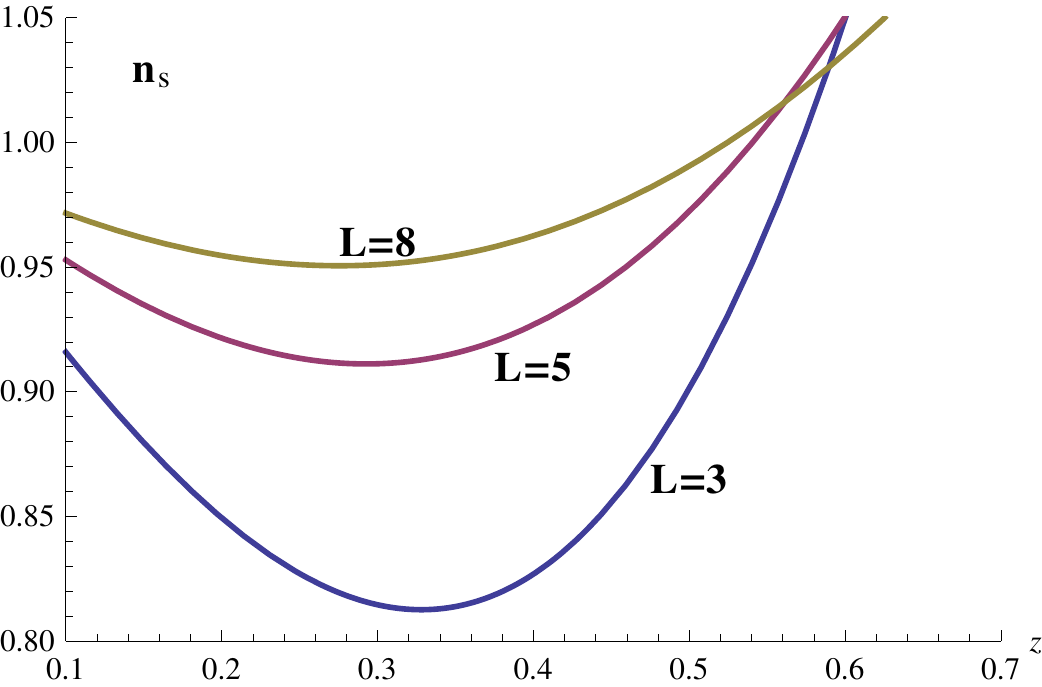}
\end{center}
\caption[]{$r$ ( left panel) and $n_s$ ( right panel ) as functions of $z$.}
\label{rfig}
 \end{figure}

Finally, in Fig. \ref{rfig} we have plotted $r$ and $n_s$ as functions of $z$ for selected values of the parameter $L$. 
One observes that for $z \sim 0.55$ large values of $r \sim 0.2$ and $n_s$ within the allowed experimental limits can be obtained. For larger values of $L$ we cannot reconcile values of $r$ in the aforementioned range with values of $n_s$ in the experimentally accepted range. In this range of $z$ and for the selected value of $z_1 = 0.9$ it is difficult to get acceptable values for 
$N \approx 55 $, as is evident from  the left pane of  Fig. \ref{efoldings}. 
Agreement with all data is achieved if we take $z_1$ very close to unity,  as shown for instance in the right panel of Fig.  \ref{efoldings}. 
 As a sample value, again taking $z_1$ to be quite close to unity, we have $ N \simeq 59$ at $z_* \simeq 0.55$, when $L \approx 3.2$, and at this point
$ r \simeq 0.24$ and $n_s \simeq 0.97$.

\section{BRIEF SUMMARY AND CONCLUSIONS}

Conformal invariance seems to be an appropriate framework as a starting point in the study of gravitation since masses of matter fields could be neglected in the regime of high curvature. Nevertheless, conformal invariance is broken at the quantum level. This breaking, referred to as the conformal anomaly, generates a local $R^2$ term in the action. In Starobinsky's view it is the quantum corrections, substantiated through this term, that give rise to inflation. The description of Starobinsky's model in terms of a scalar field gives an exponential potential with generic inflationary behavior. It is legitimate to ask whether extra matter, coupled to gravitation in a conformally invariant fashion, will modify the inflationary behavior of Starobinsky's model. 

In the present article we started with a conformally invariant action of gravitation and a scalar field incorporating the breaking of conformal invariance in the $R^2$ term. Conformal invariance allows for two distinct versions of the scalar field coupling encoded in the sign of a parameter. First we analyzed the version of the theory distinct from the Starobinsky model, which in the Einstein frame is reduced to a theory of two canonical scalar fields. No inflationary behavior can be associated with this model. Next, we turned to the particular version that includes the Starobinsky model, in which the scalar field enters as a ghost. After conformal gauge-fixing this model reduces in the Einstein frame into a theory of one scalar practically identical to the standard Starobinsky model. Inflation is generic to this model, although its quantitative signature is challenged by BICEP2. We next proceeded to introduce an additional scalar field $\sigma$, coupled in a conformally invariant fashion. Our aim was to investigate whether this inflationary behavior is affected or its quantitative profile modified. We chose to restrict our investigation by imposing a shift symmetry, i.e. invariance under shifts $\sigma\rightarrow\,\sigma+const. \;$  We found that the resulting two scalar field model possesses a class of inflationary solutions that is an attractor in field space. Furthermore, we presented an analysis of this system in order to argue that the model is essentially a one-field inflationary model corresponding to the potential $V(\phi)\,=\,\frac{1}{8\alpha}\left(1-2e^{-\sqrt{\frac{2}{3}}\phi}\,+\,L\,e^{-2\sqrt{\frac{2}{3}}\phi}\right)$. We established that slow-roll inflation of the appropriate amount occurs in this model. 
Furthermore, agreement with existing data, including the desired values of the tensor to scalar ratio can be achieved for appropriate values of the relevant parameters.

\vspace*{6mm}
\noindent
{\textbf{Acknowledgements}}

This research has been cofinanced by the European Union (European Social Fund - ESF)
and Greek national funds through the Operational Program €œEducation and Lifelong
Learning€ of the National Strategic Reference Framework (NSRF) - Research Funding
Program: {\textit{THALES-Investing in the society of knowledge through the European Social Fund.}}  K.T. would also like to thank CERN Theory Division for the hospitality
and I. Antoniadis for discussions.

\end{document}